# Spin Nomenclature for Semiconductors and Magnetic Metals

B.T. Jonker[1a)], A.T. Hanbicki[1], D.T. Pierce[2], and M.D. Stiles[2]
1) Naval Research Laboratory, Washington, D.C. 20375-5343
2) National Institute of Standards and Technology, Gaithersburg, MD 20899-8412
a) Electronic mail: Jonker@nrl.navy.mil

The different conventions used in the semiconductor and magnetic metals communities can cause confusion in the context of spin polarization and transport in simple heterostructures. In semiconductors, terminology is based on the orientation of the electron spin, while in magnetic metals it is based on the orientation of the moment. In the rapidly expanding field of spintronics, where both semiconductors and metallic metals are important, some commonly used terms ("spin-up," "majority spin") can have different meanings. Here, we clarify nomenclature relevant to spin transport and optical polarization by relating the common physical observables and "definitions" of spin polarization to the fundamental concept of conservation of angular momentum within a well-defined reference frame.

A recent Letter[1] and subsequent Comments[2,3] describing electrical spin injection from a ferromagnetic metal into a semiconductor highlight the fact that there is some confusion relating the sign of the polarization of the spin density in a semiconductor to the direction of magnetization in a ferromagnet. Confusion arises from differences in convention and nomenclature between the magnetic metals and semiconductor communities, and because of imprecise terminology. The nomenclature in the semiconductor community is based on the orientation of the carrier *spin*, an experimental observable which is accessible using standard optical spectroscopic methods. In the magnetic metals community, however, the nomenclature is based on the orientation of the electron magnetic *moment*, the basic experimental observable in a wide variety of magnetometry techniques. Since the electron spin and magnetic moment in metals are antiparallel,[4] terms such as "spin up" or "majority spin" may lead to some ambiguity and confusion – in the magnetic metals community, these terms are synonymous, and refer to an electron with a *moment parallel* to the magnetization and spin antiparallel to the magnetization.[5] The purpose of this communication is to clarify the usage of these terms. This clarification is especially timely as semiconductors come to play an increasingly important role in the rapidly expanding field of spintronics or magneto-electronics.[6]

The term "spin polarization" means definite things to many people, and frequently those definite meanings differ. Generally, polarization is the degree to which one spin state is occupied relative to the opposite spin. The connection between a measured polarization and materials properties depends on the particular measurement.[7,8] For instance, polarization can be defined for densities $P_n = (n_\uparrow - n_\downarrow)/(n_\uparrow + n_\downarrow)$, or currents $P_j = (j_\uparrow - j_\downarrow)/(j_\uparrow + j_\downarrow)$. These two quantities are not directly related and need not have the same sign or even be collinear when the appropriate vector quantities are considered. Relating them requires a transport calculation; e.g. for diffusive transport in a non-magnetic material, $(j_\uparrow - j_\downarrow) = -D\nabla(n_\uparrow - n_\downarrow)$, where $D$ is the diffusion constant. No single value of polarization describes the spin-dependent properties for a particular material. Further, in different systems, the *sign* of the polarization is determined by different conventions. In this note, we focus on these conventions. Although the material presented is basic and available in various references, given the subtleties and potential for sign errors, we hope that bringing the information together in one place will be useful.

To focus this discussion we will use as an example our experiment described in a recent Letter[1]. In this experiment we utilized the optical emission from an AlGaAs/GaAs-based quantum well light emitting diode (LED)[9] to measure the spin polarization of electrons in the quantum well which were electrically injected from a reverse biased Fe Schottky contact. The experiment was done in the Faraday geometry, shown in Fig. 1, where the applied magnetic field axis is parallel to the light propagation axis. The circular polarization of the electroluminescence, $P_{circ}$, was used to determine the orientation and magnitude of the net electron spin population of the free exciton state in the GaAs quantum well via the quantum selection rules which govern radiative recombination. These selection rules are illustrated in Figure 2.[10,11] In this experiment, only transitions involving the heavy hole ($m_j = \pm 3/2$) states participate.

For clarity, we consider the specific case where the Fe magnetization and the outgoing light are along the surface normal (out-of-plane), taken to be the +z axis. This defines a clear reference frame. In this case, the EL was dominated by positive helicity light (σ+) due to radiative recombination of $m_j = -1/2$ electrons and $m_j = -3/2$ heavy holes.[1] Note that some other experiments measure the light emitted in the reverse direction, i.e. out the back side of the wafer, and we will point out how that situation differs. Because angular momentum is conserved, it is instructive to discuss the experiment in terms of the angular momentum along the z-axis. We consider in turn the light, the spins in the semiconductor, and the spins in the ferromagnet.

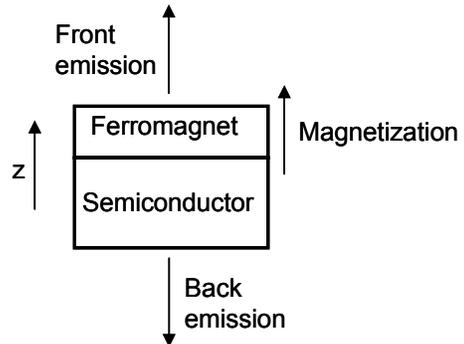

Figure 1. Geometry and directions.

The circular polarization of light is defined as

$$P_{circ} = \frac{I(\sigma^+) - I(\sigma^-)}{I(\sigma^+) + I(\sigma^-)} \quad (1)$$

where $I(\sigma^+)$ is the intensity of the light when analyzed for positive helicity. For outgoing light along the z-axis (Fig. 1), $\sigma^+ (\sigma^-)$ light has angular momentum +1 (-1) (Fig. 2). Note that different conventions call $\sigma^+$ light "left" (optics convention) or "right" circularly polarized (angular momentum convention).[12] To avoid confusion, we use here the positive / negative helicity notation because positive (negative) helicity light has angular momentum parallel (antiparallel) to the propagation direction regardless of the nomenclature used.

Before we discuss the angular momentum of the electrons in the semiconductor and ferromagnet, it is important to mention the relationship between an electron's spin and its moment. For a free electron, the spin and the moment are in opposite directions. Explicitly, they are related by $\vec{\mu} = -g\mu_B \vec{s}/\hbar$, where $|\vec{s}| = \hbar/2$ and we take $g$ to be positive for a free electron, as is customary in solid state physics.[4] It is important to note that in the most recent compilation of fundamental constants,[13] and in some texts in other fields of physics, the minus sign is absorbed in the g factor and the free electron $g$ factor is *negative*.

In the semiconductor, the electron states are designated by their total angular momentum quantum number $m_j$ relative to the quantization axis, taken here to be the surface normal.[10] The electron states at the conduction band minimum and heavy hole states at the valence band maximum are shown in Fig. 2. For simplicity, we neglect the Zeeman splitting of these states, which is small for the magnetic fields used in this experiment due to the small g-factor of GaAs ($g = -0.44$).[14] Figure 2 also indicates the two transitions relevant to the present geometry. The solid line shows the -1/2 to -3/2 transition with $\Delta m_j = -1$. To conserve angular momentum, the photon generated must take away angular momentum +1 and hence is $\sigma^+$ light when propagating in the +z direction. The converse is true for the transition shown by the dashed line. In the conduction band, the spin quantum number $m_s$ is identical to the total angular momentum quantum number $m_j$. Thus, the spin polarization in the conduction band at the time of the optical transitions is given by

$$P_s = \frac{n_{s\uparrow} - n_{s\downarrow}}{n_{s\uparrow} + n_{s\downarrow}} \quad (2)$$

where $n_{s\uparrow}$ and $n_{s\downarrow}$ refer to the number of electrons whose **spins** are parallel ($m_j = +1/2$) and antiparallel ($m_j = -1/2$) to +z. Note that by inspection of Fig. 2, we have

$$P_{circ} = -P_s \quad (3)$$

for surface emission in the Faraday geometry.

In the ferromagnet, the magnetization (or magnetic moment density) is a well defined and experimentally accessible quantity which establishes a clear reference direction. A net magnetization exists because, when integrated over filled states, there are more electrons with moments pointing in a given direction than electrons with moments pointing in the opposite direction. In the ferromagnetic metals community, these are referred to as "majority spin" or "spin up" electrons, and "minority spin" or "spin down" electrons, respectively, by historical convention.[5,15] The $g$ factors for electrons in the ferromagnetic metals Fe, Co and Ni, where there is little orbital contribution to the magnetic moment, are +2.10, 2.18, and 2.21, respectively.[4] This means the electron spin and magnetic moment for these materials are *antiparallel* (Fig. 2). Therefore, it is essential to note that the terms "majority spin" and "spin-up" are used synonymously in the magnetic metals community and refer to electrons with **moment** parallel (and *spin* antiparallel) to the magnetization.

There are many different polarizations used to describe ferromagnets. These are conventionally defined as the difference in the majority and minority values of some quantity divided by the sum. The polarization relevant to the bulk magnetization and most closely related to that used above to describe semiconductors is the polarization of the density

$$P_m = \frac{n_{maj} - n_{min}}{n_{maj} + n_{min}} \quad (4)$$

where $n_{maj}$ ($n_{min}$) refers to the number of majority (minority) spin electrons integrated over the filled states. A second, more commonly discussed polarization is defined in terms of the spin-dependent Fermi level density of states, which may differ from $P_m$ in both magnitude and sign. A third, the polarization of the conductivity, describes transport in bulk ferromagnets. For the experiment of interest here, a fourth, the polarization of the interface conductance is most relevant. While the polarization of the bulk density is positive by definition, the other polarizations need not be. For Fe, the polarization of the Fermi level density of states is positive, and the polarization of the conductivity is negative. In contrast, for Co and Ni, the polarization of the Fermi level density of states is negative, and the polarization of the conductivity is positive.

|  | Moment along z | Angular momentum along z | Helicity |
|---|---|---|---|
| **Light** | | | |
| Front emission along z | | +1 | $\sigma^+$ |
| | | -1 | $\sigma^-$ |
| Back emission along -z | | +1 | $\sigma^-$ |
| | | -1 | $\sigma^+$ |
| **Ferromagnet** | | | |
| Majority | $g\mu_B/2$ | -1/2 | |
| Minority | $-g\mu_B/2$ | 1/2 | |
| **Semiconductor** | | | |

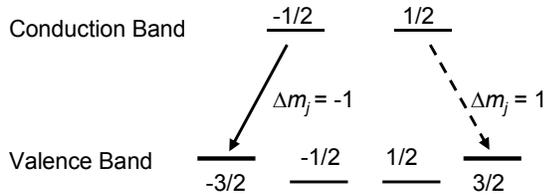

Figure 2. Angular momenta for light emission from GaAs. The third column shows the angular momenta (in units of $\hbar$) of light and electrons in the ferromagnet and semiconductor. The second column gives the moment of the same electrons in the ferromagnet and the final column gives the helicity of the light. At the bottom, the third column is expanded to give the angular momentum of states in the semiconductor at the conduction band minimum and valence band maximum, and illustrates the relevant heavy hole transitions.

Returning to our example of the experiment of Ref. 1, measuring positive helicity light for surface emission in the Faraday geometry implies that there are a greater number of $m_j = -1/2$ electrons than $m_j = +1/2$ electrons in the conduction band (free exciton state) of the GaAs quantum well. The $m_j = -1/2$ electrons have *spins* that are parallel to those of the majority spin electrons in the Fe contact. Thus majority spins have been preferentially injected from the reverse biased Fe Schottky tunnel contact into the semiconductor. The observation that the tunneling current is dominated by Fe majority spin carriers is consistent with the model proposed by Stearns[16] and the tunneling experiments of Meservey and Tedrow.[5] However, the measured electron polarization in the quantum well is not trivially related to the magnitude of the polarization of the current crossing the barrier. This value must be inferred via a model dependent transport calculation.

If the circular polarization of the light is measured out the back side of the structure a similar analysis obviously applies. If the light is collected in the reverse direction, one would expect to find light of the opposite helicity as is shown in Fig. 2. To keep track of signs and to enable comparison of results of different experiments, the quantization axis chosen for the semiconductor spins, and how other quantities like magnetic field are referenced to it, should be explicitly stated. If the quantization direction is chosen to be the light propagation direction, for instance, the transitions in the semiconductor have the same form regardless of whether light comes out the front or back.

In summary, the nomenclature in the semiconductor community is based on the orientation of the carrier *spin*, an experimental observable which is accessible using standard optical spectroscopic methods, while in the magnetic metals community the nomenclature is based on the orientation of the *moment*, the basic experimental observable in a wide variety of magnetometry techniques. The respective terminology which is commonly employed can result in some confusion as these fields interact in the semiconductor spintronics arena. Consequently, care must be taken to insure accurate communication of experimental and theoretical results. Confusion can be avoided by describing everything in terms of angular momenta in a common reference frame.

This work was supported in part by the Office of Naval Research and the DARPA SpinS program.